\documentclass[twocolumn,aps]{revtex4}
\usepackage{amssymb}
\newcommand{\half}{\frac{1}{2}}
\newcommand{\thrd}{\frac{1}{3}}
\newcommand{\frth}{\frac{1}{4}}
\newcommand{\sxth}{\frac{1}{6}}
\begin{document}
\title{Conformal Gravity: Dark Matter and Dark Energy}
\author{Robert K. Nesbet }
\affiliation  {IBM Almaden Research Center,
650 Harry Road, San Jose, CA 95120, USA }
\begin{abstract}
This short review examines recent progress in understanding
dark matter, dark energy, and galactic halos using theory that 
departs minimally from standard particle physics and cosmology.
Strict conformal symmetry (local Weyl scaling covariance), postulated
for all elementary massless fields, retains standard fermion and
gauge boson theory but modifies Einstein-Hilbert 
general relativity and the Higgs scalar field model, with no
new physical fields.  Subgalactic phenomenology is retained.  Without
invoking dark matter, conformal gravity and a conformal Higgs model 
fit empirical data on galactic rotational velocities, galactic halos,
and Hubble expansion including dark energy.
\end{abstract}

\maketitle
\section{Introduction}
\par The current consensus paradigm for cosmology is the $\Lambda$CDM
model\cite{DOD03}.  Here $\Lambda$ refers to dark energy, whose 
existence is inferred from accelerating Hubble expansion of the cosmos, 
while CDM refers to cold dark matter, observable to date only through
its gravitational effects.  The underlying assumption is that 
general relativity, as originally formulated by Einstein and verified
by observations in our solar system, is correct without modification on 
the vastly larger scale of galaxies. Extrapolating back in time, initial
big-bang cosmic inflation is an independent postulate.
\par Dark energy, dark matter, and the big-bang concept are reconciled
only with some difficulty to some of the principles deduced from
traditional laboratory and terrestrial physics.  In particular, it is 
not obvious that traditional thermodynamics can be assumed for studying
extreme situations such as cosmic inflation and the collapse of matter 
into black holes.  
\par In the interest of reducing such uncertainties, the present review
considers recent evidence supporting a theory, with minimal deviation 
from well-established theory of fields and particles,  that fits 
the same cosmological data that motivates $\Lambda$CDM, while explaining
dark energy, motivating early cosmic expansion, and removing the need 
for dark matter.  
\par In the theory considered here the simple postulate of universal
conformal symmetry for all elementary (massless) fields combines
conformal gravity\cite{WEY18,MAN90,MAN06} with a conformal scalar field
model\cite{MAN90}, introducing no new fundamental fields\cite{NESM1}.
While other modified gravitational theories have been shown to account
for aspects of empirical cosmology, the postulate of universal conformal
symmetry proposed here is a minimalist baseline requiring extension
only if found to be in conflict with observation.
\par Accepted theories of massless fermion and gauge boson fields
exhibit strict conformal symmetry\cite{DEW64}, defined by invariance
of an action integral under local Weyl scaling\cite{WEY18}, such that
$g_{\mu\nu}(x)\to g_{\mu\nu}(x)\Omega^2(x)$ for fixed coordinate values.
For a scalar field, $\Phi(x)\to\Phi(x)\Omega^{-1}(x)$.  Standard general
relativity and the electroweak Higgs model are not conformal. Since a
conformal energy-momentum tensor must be traceless, this suggests a
fundamental inconsistency\cite{MAN90,MAN06}.  The gravitational field
equation equates the Einstein tensor, with nonvanishing trace, to
the traceless tensors of quantum fields.  Dynamical interactions which
produce elementary particle mass redistribute energy terms among the
interacting fields, while the total energy-momentum tensor remains
traceless\cite{MAN11}.  In the conformal Higgs model, a dynamical term
breaks symmetry and determines dark energy while preserving the
conformal trace condition\cite{NESM1,NESM2}.
\par Postulating conformal symmetry for all elementary fields modifies
both gravitational and electroweak theory\cite{MAN90,NESM1}.
Conformal gravity\cite{MAN06} retains the logical structure
of general relativity, but replaces the Einstein-Hilbert Lagrangian
density, proportional to the Ricci curvature scalar, by a quadratic
contraction of the conformal Weyl tensor\cite{WEY18}.  This removes
the inconsistency of the gravitational field equation.  Mannheim and
Kazanas\cite{MAK89} showed that this preserves subgalactic
phenomenology, modifying gravitation only on a galactic scale.
Formal objections to this conclusion\cite{FLA06} have been refuted in
detail by Mannheim\cite{MAN07}.
\par Conformal theory, not invoking dark matter, was shown some time ago
to fit observed excessive rotation velocities outside galactic cores for
eleven typical galaxies, using only two universal
constants\cite{MAN97,MAN06}.
More recently, rotation velocities for 138 dwarf and spiral galaxies
whose orbital velocities are known outside the optical disk have been
fitted to conformal gravity\cite{MAO11,MAO12,OAM12}.
The data determine a third parameter that counteracts an otherwise
increasing velocity at very large radii.
\par The Higgs mechanism for gauge boson mass determines a nonvanishing
scalar field amplitude that breaks both conformal and SU(2) gauge
symmetries.  The Higgs mechanism is preserved in conformal theory, but
the tachyonic mass parameter $w^2$ of the Higgs model is required to be
of dynamical origin. In uniform, isotropic geometry, conformal
gravitational and Higgs scalar fields imply a modified Friedmann cosmic
evolution equation\cite{NESM1}.  Parameter $w^2$ determines a
cosmological constant (dark energy)\cite{NESM2}.
\par The modified Friedmann equation has been parametrized to fit
relevant cosmological data within empirical error limits, including 
dark energy but not invoking dark matter\cite{NESM1}.  The integrated
Friedmann scale parameter indicates that mass and radiation density 
drive cosmic expansion in the early universe, while cosmic acceleration 
is always positive.  The cosmological time dependence of nominally 
constant parameters of the conformal Higgs model couples scalar and 
gauge fields and determines parameter $w^2$. The implied cosmological 
constant, an unanticipated consequence of the standard model Higgs
mechanism, is in order-of-magnitude agreement with its empirical 
$\Lambda$CDM value\cite{NESM1,NESM2}.
\par Conformal theory is consistent with a model of galactic halos
that does not require unobservable dark matter\cite{NESM4}.  Hence 
conformal theory removes the need for dark matter except possibly for 
galactic clusters.  As shown below, the postulate of universal conformal
symmetry significantly alters theory relevant to galaxy and cluster 
formation.  The implications have not yet been incorporated into a
dynamical model.

\section{Postulates and formalism}
\par Conformal gravity theory has recently been reviewed by
Mannheim\cite{MAN06}.  Conventions used by Mannheim are modified here in
some details to agree with electroweak theory references, in particular
as applied to the Higgs scalar field\cite{PAS95,CAG98}.  Sign changes 
can arise from the use here of flat-space diagonal metric 
$\{1,-1,-1,-1\}$ for contravariant coordinates $x^\mu=\{t,x,y,z\}$. 
Natural units are assumed with $c=\hbar=1$.
\par Variational theory for fields in general relativity is a
straightforward generalization of classical field theory\cite{NES03}.
Given Riemannian scalar Lagrangian density ${\cal L}$, action integral
$I=\int d^4x \sqrt{-g} {\cal L}$ is required to be stationary for
all differentiable field variations, subject to appropriate boundary
conditions.  The determinant of metric tensor $g_{\mu\nu}$ is denoted
here by $g$.  Gravitational field equations are determined by metric
functional derivative
$X^{\mu\nu}= \frac{1}{\sqrt{-g}}\frac{\delta I}{\delta g_{\mu\nu}}$.
Any scalar ${\cal L}_a$ determines energy-momentum tensor
$\Theta_a^{\mu\nu}=-2X_a^{\mu\nu}$,
evaluated for a solution of the coupled field equations.
Generalized Einstein equation $\sum_aX_a^{\mu\nu}=0$ is expressed as
$X_g^{\mu\nu}=\half\sum_{a\neq g}\Theta_a^{\mu\nu}$.  Hence summed trace
$\sum_ag_{\mu\nu}X_a^{\mu\nu}$ vanishes for exact field solutions. Trace
$g_{\mu\nu}X_a^{\mu\nu}=0$ for a bare conformal field\cite{MAN06}.
\par Weyl tensor $C_\lambda^{\mu\kappa\nu}$, a traceless projection of 
the Riemann tensor \cite{WEY18,MAN06}, defines a conformally invariant
action integral, with Lagrangian density
${\cal L}_W=-\alpha_g C_\lambda^{\mu\kappa\nu} 
  C^\lambda_{\mu\kappa\nu}$.
Removing a 4-divergence\cite{MAN06},
\begin{eqnarray}\label{Lg} 
{\cal L}_g=-2\alpha_g( R^{\mu\nu}R_{\mu\nu}-\thrd R^2 ).
\end{eqnarray}
Here Ricci tensor $R^{\mu\nu}$, 
a symmetric contraction of the Riemann tensor, defines
Ricci scalar $R=g_{\mu\nu}R^{\mu\nu}$.
The relative coefficient of the two quadratic terms in ${\cal L}_g$
is fixed by conformal symmetry\cite{MAN06}.
\par The metric tensor in quadratic line element
$ds^2=g_{\mu\nu}dx^\mu dx^\nu$ is determined by gravitational field
equations.  Outside a bounded spherical source density, the field
equations implied by conformal ${\cal L}_g$ have an exact
solution\cite{MAK89} given by static exterior Schwarzschild (ES) metric
\begin{eqnarray}\label{ES}
ds^2_{ES}=B(r)dt^2-\frac{dr^2}{B(r)}-r^2d\omega^2, 
\end{eqnarray}
where $d\omega^2=d\theta^2+\sin^2\theta d\phi^2$.
Gravitational potential
$B(r)=1-2\beta/r+\gamma r-\kappa r^2$,
with constants of integration $\beta,\gamma,\kappa$.   
These constants extend Birkhoff's theorem\cite{DAF05}, which implies
constant $\beta$ for standard general relativity, to conformal gravity.
\par A uniform, isotropic cosmos with Hubble expansion is described
by the Robertson-Walker (RW) metric
\begin{eqnarray}\label{RW}
ds^2_{RW}=dt^2-a^2(t)(\frac{dr^2}{1-kr^2}+r^2d\omega^2),
\end{eqnarray} 
where $k$ is a curvature constant.
\par A conformally invariant action integral is defined for complex
scalar field $\Phi$ by Lagrangian density\cite{MAN06,NESM1,NESM2} 
\begin{eqnarray}\label{LPhi}
{\cal L}_\Phi=
 (\partial_\mu\Phi)^\dag\partial^\mu\Phi-\sxth R\Phi^\dag\Phi
 -\lambda(\Phi^\dag\Phi)^2,
\end{eqnarray}
where $R$ is the Ricci scalar.  The Higgs mechanism\cite{CAG98}  
postulates incremental Lagrangian density 
$\Delta{\cal L}_\Phi=w^2\Phi^\dag\Phi-\lambda(\Phi^\dag\Phi)^2$,
replacing $-\lambda(\Phi^\dag\Phi)^2$.
Because term $w^2\Phi^\dag\Phi$ breaks conformal symmetry, universal 
conformal symmetry requires it to be produced dynamically.  
The scalar field equation, including $\Delta{\cal L}_\Phi$, is 
\begin{eqnarray}\label{eqPhi}
\partial_\mu\partial^\mu\Phi= 
 (-\sxth R+w^2-2\lambda\Phi^\dag\Phi)\Phi.
\end{eqnarray}
If derivatives in $\partial_\mu\partial^\mu\Phi$ can be neglected,
this equation has an exact solution
$\Phi^\dag\Phi=\phi_0^2=(w^2-\sxth R)/2\lambda$
\cite{NESM1,CAG98}.
\par Cosmic Hubble expansion is conventionally considered in uniform,
isotropic geometry, using the Robertson-Walker metric, for which the 
conformal action integral of Weyl vanishes identically\cite{MAN06}.
Conformal gravity is inactive in this context, but a conformal scalar 
field affects gravity through the Ricci scalar term in its Lagrangian 
density.  In conformal theory, Hubble expansion is determined by a 
scalar field\cite{MAN06}.  Since a Higgs scalar field must exist, to 
produce gauge boson masses in electroweak theory, the simplest way to
account for both established cosmology and electroweak physics is to
equate the cosmological and Higgs scalar fields\cite{NESM2}.  
Postulating universal conformal symmetry, this requires no otherwise
unknown fields or particles.

\section{Dark matter: galactic rotation velocities}
\par The concept of dark matter originates from dynamical
studies\cite{OAP73,OPY74} which indicate that a spiral galaxy
such as our own would lack long-term stability if not augmented by an
additional gravitational field from some unseen source.  This concept 
is supported by other cosmological data\cite{SAN10}.  Anomalous excess 
centripetal acceleration, observed in orbital rotation velocities
\cite{RAS72,BOS81} and gravitational lensing\cite{WCW79,PAC87,BAR10},
is attributed to a dark matter galactic halo\cite{PSS98,SAL07}. 
The parametrized Friedmann cosmic expansion equation of standard theory 
requires a large dark matter mass density\cite{SAN10}.
A central conclusion of standard $\Lambda$CDM cosmology is that the 
inferred dark matter significantly outweighs observed baryonic 
matter\cite{DOD03,SAN10}.
\par In standard $\Lambda$CDM theory\cite{DOD03}, phenomena and data
that appear to conflict with Einstein general relativity are attributed
to dark matter\cite{SAN10}, assumed to be essentially unobservable 
because of negligible direct interaction with radiation or baryonic 
matter.  An important logical point is that if dark matter is identified
only by its gravitational field, it is not really an independent entity.
Any otherwise unexplained gravitational field, entered into Poisson's 
equation, implies a source density, which can conveniently be labelled 
as dark matter.  Attributing physical properties to dark matter, other 
than this pragmatic definition as a field source, may be an empty 
exercise.
\par An alternative strategy is to treat non-Einsteinian phenomena as
evidence for failure or inadequacy of the theory.  The MOND (modified 
newtonian dynamics) model of Milgrom\cite{MIL83}, motivated by anomalous
velocities $v$ observed for dust or hydrogen gas in outer galactic
circular orbits, has been very successful in fitting empirical
data\cite{SAM02,SAN10}.  Observed velocities are constant or increasing
at large radius $r$, while Keplerian $v^2$ would drop off as $1/r$.
MOND models this effect by modifying Newton's second law for low 
acceleration $a\leq a_0$, a universal constant not defined
by standard relativity.  
\par MOND replaces acceleration $a$ by $a\mu(a/a_0)\to a^2/a_0$ as 
$a/a_0\to0$ in Newton's law $F/m=a$.  For a Keplerian circular orbit, 
$F/m=GM/r^2$.  Then $a=v^2/r \to v^4/(a_0r^2)$ implies $v^4=a_0GM$, 
explicitly the Tully-Fisher (TF) relation\cite{TAF77}, where $M$ is 
galactic baryonic mass\cite{MCG11}.  The empirical TF relation is 
inferred from observed galactic orbital rotation velocities\cite{SAN10}.
\par This empirical $v^4$ law appears to be valid in particular for
largely gaseous galaxies\cite{MCG11}, whose baryonic mass is 
well-defined.  It does not follow readily from the $\Lambda$CDM
model of a dark matter galactic halo\cite{SAM02,SAN10}.
Gravitational theory can be revised specifically to agree with
TF\cite{BEK04}, at the cost of postulating otherwise unknown scalar and 
vector fields, but this does not necessarily describe other phenomena
such as Hubble expansion.  The relativistic theory of Moffat\cite{MOF06}
has been parametrized to fit rotation velocities for a large set of
galaxies\cite{BAM06}.  A massive vector field is introduced and nominal
constants are treated as variable scalar fields. The theory describes
other aspects of cosmology including gravitational lensing\cite{MRT12}.
\par In conformal theory\cite{MAK89} the most general 
spherically symmetric static exterior Schwarzschild metric 
outside a source density defines a relativistic gravitational potential 
$B(r)=1 -2\beta/r +\gamma r -\kappa r^2$.  
A circular orbit with velocity $v$ is stable if
$v^2=\half rdB/dr=\beta/r +\gamma r/2 -\kappa r^2$.
If dark matter is omitted, parameter $\beta=GM$, proportional to total
galactic baryonic mass $M$.  Defining $N^*$ as total visible plus 
gaseous mass in solar units, and neglecting $\kappa$,
Mannheim\cite{MAN97} determined two universal parameters such that
$\gamma=\gamma^*N^*+\gamma_0$ fits rotational data for eleven typical 
galaxies, not invoking dark matter\cite{MAN97,MAN06}. Parameter 
$\gamma_0$, independent of galactic mass, implies an isotropic 
cosmological source.  Hence the parametrized gravitational field forms a
spherical halo. A consistency test, if adequate data are available, is 
that the same field should account for gravitational lensing.  
\par Constant of integration $\kappa$ determines a radius at which 
incremental radial acceleration vanishes. This removes the objection
that $\gamma$ by itself would imply indefinitely increasing velocities.
The fit of conformal gravity to rotational data\cite{MAN97} has recently
been extended, including parameter $\kappa$, to 111 spiral galaxies 
whose orbital velocities are known outside the optical disk\cite{MAO11}.
$\kappa$ is treated as a global constant, not dependent on mass or on a
specific boundary condition.
\par The fit of mass-independent $\gamma_0$ to observed data implies a 
significant effect of the cosmic background, external to a baryonic 
galactic core\cite{MAN97}. Parameter $\gamma_0$ is equivalent to cosmic 
background curvature\cite{MAK89,MAN06}.  Attributed to a galactic 
halo, this is a direct measurement of a centripetal effect. 
In the conformal halo model\cite{NESM4}, discussed below,
total galactic mass $M$ determines halo radius $r_H$,
so that $\kappa_H=\gamma_0/2r_H$ is a function of $M$. 
It would be informative to fit rotation data 
using parameters $\kappa_H$ and explicitly mass-dependent
$\kappa^*N^*$.  The conformal halo model is consistent with 
conformal theory of both anomalous rotation and the Hubble 
expansion\cite{NESM1}.  
\par  Conformal gravity has been shown to be consistent with the TF
relation\cite{MAN06}. This argument is supported by the conformal model
of galactic halos, described below\cite{NESM4}.
Outside the galactic core, but for $r\ll r_H$,
conformal velocity function $v^2(r)=GM/r+\gamma r/2$
has a broad local minimum at $r_x^2=2GM/\gamma$.
Evaluated at $r_x$, $\gamma r_x/2=GM/r_x$, such that
$v^4(r_x)=4(\gamma r_x/2)(GM/r_x)=2\gamma GM$.
If $\gamma^*N^*\ll\gamma_0$ and dark matter is omitted, this is an
exact baryonic Tully-Fisher relation, as inferred from recent analysis
of galactic data\cite{MCG11}.  Centripetal acceleration
at $r_x$ determines MOND parameter\cite{SAN10} $a_0=2\gamma$.
 
\section{Dark matter: Hubble expansion}
\par It was first recognized by Hubble\cite{HUB29} that galaxies visible
from our own exhibit a very regular centrifugal motion, characterized
as uniform expansion of the cosmos\cite{DOD03}.  Redshift $z$, a 
measure of relative velocity, is nearly proportional to a measure 
of distance deduced from observed luminosity.
Refining the observed data by selecting Type Ia 
supernovae as "standard candles", cosmic expansion has been found to be 
accelerating in the current epoch\cite{RAS98,PER99}.
\par That Einstein's equations can imply expansion of a uniform,
isotropic universe was first shown by Friedmann\cite{FRI22,FRI24} and
LeMa\^itre\cite{LEM27}.  This is described by the Friedmann equations,
which determine cosmic scale parameter $a(t)$ and deceleration
parameter $q(t)$.
\par In conformal gravitational theory the Einstein-Hilbert Lagrangian 
density is replaced by a uniquely determined quadratic contraction of 
the Weyl tensor, which vanishes identically in uniform, isotropic 
Robertson-Walker (RW) geometry.  Vanishing of the metric functional 
derivative of action integral $I_g$\cite{MAN06}, for the RW metric, 
can be verified by direct evaluation. 
\par  The conformal gravitational action integral replaces the 
standard Einstein-Hilbert action integral, but in the uniform
model of cosmology its functional derivative drops out completely from
the gravitational field equations.  The observed Hubble expansion 
requires an alternative gravitational mechanism.  This is supplied by a 
postulated conformal scalar field\cite{MAN06}.  A nonvanishing conformal
scalar field determines gravitational field equations that differ from 
Einstein-Hilbert theory. The Newton-Einstein gravitational constant $G$ 
is not relevant.  As shown by Mannheim\cite{MAN03}, the gravitational 
constant determined by a scalar field is inherently negative. 
\par The conformal Higgs model\cite{NESM1} differs from Mannheim 
because scalar field Lagrangian terms proportional to $\Phi^\dag\Phi$
in Higgs and conformal theory have opposite algebraic signs.  
A consistent theory must include both terms and solve interacting 
gravitational and scalar field equations.  This determines a modified
field equation in which Einstein tensor $R^{\mu\nu}-\half g^{\mu\nu}R$,
where $R^{\mu\nu}$ is the Ricci tensor and $R= g_{\mu\nu}R^{\mu\nu}$,
is replaced by tensor $R^{\mu\nu}-\frth g^{\mu\nu}R$, traceless as 
required by conformal theory\cite{MAN06}. In uniform RW geometry  
this determines a modified Friedmann evolution equation\cite{NESM1} 
that differs from the standard equation used in all previous work, 
including that of Mannheim\cite{MAN03}.
\par In the standard Einstein equation, 
$R^{\mu\nu}-\half Rg^{\mu\nu}+\Lambda g^{\mu\nu}=
 -8\pi G\Theta_m^{\mu\nu}$, $\Theta_m^{\mu\nu}$
is the energy-momentum tensor due to matter and radiation.
Radiation energy density can be neglected in the current epoch. 
Cosmological constant $\Lambda$ must be determined empirically.
\par For uniform cosmic mass-energy density $\rho_m$, in RW geometry 
the $R^{00}$ Einstein equation reduces to standard Friedmann equation 
$\frac{{\dot a}^2}{a^2}+\frac{k}{a^2}=\thrd(\kappa\rho_m+\Lambda)$. Here
${\dot a}/a=h(t)$ is defined in Hubble units such that at present time
$t_0$, $h(t_0)=1$, $a(t_0)=1$ and coefficient $\kappa=8\pi G$. 
This implies sum rule $\Omega_m+\Omega_k+\Omega_\Lambda=1$,
usually presented as a pie-chart for the energy budget of the universe.
The dimensionless weight functions are
$\Omega_m(t)=\frac{\kappa\rho_m(t)}{3h^2(t)}, 
 \Omega_k(t)=-\frac{k}{a^2(t)h^2(t)},
 \Omega_\Lambda(t)=\frac{\Lambda}{3h^2(t)}$.
The second Friedmann equation determines acceleration weight 
$\Omega_q=-q=\frac{{\ddot a}a}{{\dot a}^2}$.
\par In standard $\Lambda$CDM, curvature parameter 
$\Omega_k(t_0)$ is negligible while dark energy 
$\Omega_\Lambda(t_0)=0.73$ and mass
$\Omega_m(t_0)=0.27$\cite{KOM09,KOM11}.
This empirical value of $\Omega_m$ is much larger than implied by the 
verifiable density of baryonic matter, providing a strong argument
for abundant dark matter\cite{SAN10}.
\par Mannheim\cite{MAN03} showed that Type Ia supernovae data for 
redshifts $z\leq 1$ could be fitted equally well with 
$\Omega_q(t_0)=-q=0.37$ and $\Omega_\Lambda(t_0)=0.37$, 
assuming $\Omega_m=0$.  This argues against the need for dark matter.
However, for $\Omega_m=0$, the standard Friedmann sum rule reduces to
$\Omega_k+\Omega_\Lambda=1$.  This would imply current
curvature weight $\Omega_k(t_0)=0.63$, much larger than its consensus 
empirical value\cite{KOM11}. The modified Friedmann equation derived 
from conformal Higgs theory\cite{NESM1} avoids this problem.  Fitted
parameters, without dark matter, are consistent with current 
cosmological data\cite{KOM11}.  Anomalous imaginary-mass term $w^2$ in 
the Higgs scalar field Lagrangian becomes a cosmological constant (dark 
energy) in the modified Friedmann equation\cite{NESM2}.  Dark energy 
dominates the current epoch.

\section{Dark energy}
\par Coupled scalar and gauge boson fields produce gauge boson mass 
through the Higgs mechanism\cite{CAG98}, starting in the electroweak 
transition epoch.  The universal conformal symmetry postulate requires 
Higgs parameter $w^2$, which breaks conformal symmetry, to be a 
dynamical consequence of the theory. Conformal symmetry extends the 
Higgs model to include the metric tensor field\cite{NESM2}. The modified
Friedmann equation determines cosmological time variation of Ricci 
scalar $R$, present in the Lagrangian density of the bare conformal 
scalar field. This in turn induces a neutral gauge current density that 
dresses the scalar field with an induced gauge field\cite{NESM2}. 
This determines parameter $w^2$, which becomes dark energy in the 
modified Friedmann equation, preserving the Higgs mechanism for gauge 
boson masses and the trace condition for the coupled field equations. 
\par In conformal Higgs theory\cite{NESM1,NESM2}, the vanishing trace 
condition removes the second Friedmann equation and the sum rule 
becomes $\Omega_m+\Omega_k+\Omega_\Lambda+\Omega_q=1$.  
Higgs scalar field constants $\phi_0,w^2$\cite{NESM1,CAG98} define 
effective gravitational parameters ${\bar\kappa}=-3/\phi_0^2$
and ${\bar\Lambda}=\frac{3}{2}w^2$.
This results in dimensionless weight functions 
$\Omega_m(t)=\frac{2{\bar\kappa}\rho_m(t)}{3h^2(t)},
 \Omega_k(t)=-\frac{k}{a^2(t)h^2(t)},
 \Omega_\Lambda(t)=\frac{w^2}{h^2(t)},
 \Omega_q(t)=\frac{{\ddot a}(t)}{a(t)h^2(t)}$.
Solving the modified Friedmann equation with $\Omega_m=\Omega_k=0$, a 
fit to Type Ia supernovae magnitude data for redshifts $z\leq1$ finds
$\Omega_\Lambda(t_0)=0.732$\cite{NESM1}, in agreement with consensus 
empirical value $\Omega_\Lambda(t_0)=0.726\pm0.015$\cite{KOM11}. 
The computed acceleration weight is $\Omega_q(t_0)=0.268$.
Note that only one effective independent parameter is involved in 
fitting the modified Friedmann equation to $z\leq1$ redshift data.
\par Fitting conformal gravitation to galactic rotation data, the
Schwarzschild gravitational potential $B(r)$ contains a universal
nonclassical term $\gamma_0 r$\cite{MAK89}.
Coefficient $\gamma_0$, independent of galactic luminous mass, must
be attributed to the background Hubble flow\cite{MAN03}.  On converting 
the local Schwarzschild metric to conformal RW form, this produces a 
curvature parameter $k=-\frth\gamma_0^2$\cite{MAK89} which is small and 
negative, consistent with other empirical data.  This supports the 
argument for modifying the standard Friedmann equation.
\par The modified Friedmann equation determines scale parameter $a(t)$ 
and Hubble function $h(t)=\frac{{\dot a}}{a}(t)$, for redshift 
$z(t)=1/a(t)-1$.  A numerical solution from $t=0$ to current $t=t_0$ 
is determined by four fixed parameters\cite{NESM1}.  Adjusted to
fit two dimensionless ratios characterizing CMB acoustic peak structure
\cite{WAM07}, as well as $z\leq1$ Type Ia supernovae magnitudes,
implied parameter values $\Omega_a(t_0)$ to three decimals are:
$\Omega_\Lambda=0.717,\Omega_k=0.012,\Omega_m=0.000,\Omega_r=0.000$,
with computed acceleration weight $\Omega_q=0.271$\cite{NESM1}.  
Consensus empirical values are $\Omega_\Lambda=0.725\pm0.016$,
$\Omega_k=-0.002\pm0.011$\cite{KOM11}. 
\par In the current epoch, dark energy and acceleration terms are of 
comparable magnitude, the curvature term is small, and other terms are 
negligible.  The negative effective gravitational constant implies
energy-driven rapid inflation of the early universe.  Hubble function
$h(t)$ rises from zero to a maximum at $z=1371$, prior to the CMB
epoch, then descends as $t\to\infty$ to a finite asymptotic value 
determined by the cosmological constant\cite{NESM1}.
Acceleration weight ${\ddot a}a/{\dot a}^2$ is always positive.
Although deduced from the same data fitted by standard $\Lambda$CDM,
the implied behavior of the early universe is significantly different.
Whether this is consistent with a big-bang singularity at $t=0$ is
at present difficult to assess, since the time-dependence of nominally 
constant Higgs model parameters is not yet known. 
\par The standard Higgs mechanism, responsible for gauge field mass, can
be derived using classical U(1) and SU(2) gauge fields, coupled to Higgs
SU(2) doublet scalar field $\Phi$ by covariant derivatives\cite{CAG98}.
In the conformal Higgs model, dark energy occurs as a property of the
finite Higgs scalar field produced by this symmetry-breaking mechanism 
\cite{NESM1,NESM2}.
Ricci scalar $R$ in the conformal scalar field Lagrangian density 
requires extending the Higgs model to include the classical 
relativistic metric tensor.  If derivatives of $\Phi$ can be neglected,
scalar field Eq.(\ref{eqPhi}) has an exact solution given by  
\begin{eqnarray}
\Phi^\dag\Phi=\phi_0^2=(w^2-\sxth R)/2\lambda.
\end{eqnarray}
The phase is arbitrary, so $\phi_0$ can be a real constant.  Its
experimental value is $\phi_0=180GeV$\cite{CAG98}. 
Consistent with the modified Friedmann equation of conformal theory,
empirical dark energy weight $\Omega_\Lambda=w^2=0.717$\cite{NESM1},
in Hubble units.  Hence $w=0.847\hbar H_0=1.273\times 10^{-33}eV$, 
where $H_0$ is the Hubble constant.
\par In conformal theory, dark energy appears in the energy-momentum
tensor of the scalar field required by the Higgs mechanism to produce
gauge boson masses.  The implied cosmological constant can be computed
as the self-interaction of the Higgs scalar field due to induction of
an accompanying gauge boson field\cite{NESM2}. The required  transition 
amplitude depends on the cosmological time derivative of the dressed
scalar field.
\par From the scalar field equation,
$\phi_0^2=-\zeta/2\lambda$, where $\zeta=\sxth R-w^2$.
Computed from the integrated modified Friedmann equation, 
$\zeta(t_0)=1.224\times 10^{-66}eV^2$\cite{NESM2}. Given
$\phi_0=180GeV$, the empirical value of dimensionless Higgs parameter
$\lambda=-\half\zeta/\phi_0^2$ is $-0.189\times 10^{-88}$\cite{NESM2}.
For $\lambda<0$ the conformal Higgs scalar field does not have
a stable fluctuation\cite{NESM3}, required to define a massive Higgs
particle.  The recent observation of a particle or resonance at
125 GeV is consistent with such a Higgs boson, but may prove to be an
entirely new entity when more definitive secondary properties are
established\cite{CHO12}.  Because the conformal Higgs field  
retains the finite constant field amplitude essential to gauge boson
and fermion mass, while accounting for empirically established dark
energy, an alternative explanation of the recent 125GeV resonance 
might avoid a severe conflict with observed cosmology.
\par Expressed in Hubble weights for the modified Friedmann equation,
the RW metric Ricci scalar is
$R=6\frac{{\dot a}^2}{a^2}(1-\Omega_k+\Omega_q)$,
which depends on $a(t)$.
$\phi_0^2=(w^2-\sxth R)/2\lambda$ is not strictly constant, but 
varies in time on a cosmological scale ($\sim 10^{10}yrs$).  
Numerical solution of the modified Friedmann equation\cite{NESM1},
with fixed $w^2$ and $\lambda$, implies logarithmic time derivative
$\frac{{\dot\phi}_0}{\phi_0}(t_0)=-2.651 H_0$.
\par This cosmological time derivative defines a very small scale 
parameter that drives dynamical coupling of scalar and gauge fields, 
in turn determining Higgs parameter $w^2$\cite{NESM2}.  This offers an
explanation, unique to conformal theory, of the huge disparity in 
magnitude between parameters relevant to cosmological and 
elementary-particle phenomena.
\par Solving the coupled field equations for $g_{\mu\nu},\Phi$, and 
induced neutral gauge field $Z_\mu$, using computed time derivative
$\frac{{\dot\phi}_0}{\phi_0}(t_0)$, gives $w\simeq2.651\hbar H_0$
$=3.984\times 10^{-33}eV$\cite{NESM2}.  A more accurate calculation 
should include charged fields $W^\pm_\mu$ and the presently unknown 
time dependence of Higgs parameter $\lambda$. The approximate
calculation\cite{NESM2} agrees in magnitude with the value implied by
dark energy Hubble weight $\Omega_\Lambda(t_0)=0.717$:
$w=1.273\times 10^{-33}eV$.  These numbers justify the conclusion that 
conformal theory explains both existence and magnitude of dark energy.

\section{Galactic halos}
\par A galaxy forms by condensation of matter from uniform, isotropic 
background density $\rho_m$ into observed galactic density $\rho_g$. 
Conservation of mass and energy requires that total galactic mass $M$ 
must be missing from a surrounding depleted background.  Since this 
is uniform and isotropic, it can be modeled by a depleted sphere of 
radius $r_H$, such that $4\pi\rho_m r_H^3/3=M$.  In particular, the 
integral of $\rho_g-\rho_m$ must vanish.  Any gravitational 
effect due to this depleted background could be attributed to a 
spherical halo of dark matter surrounding a galaxy.  This is the 
current consensus model of galactic halos\cite{DOD03,SAN10,SAL07}.  
Conformal theory provides an alternative interpretation of observed 
effects, including lensing and anomalous galactic rotation, as 
gravitational effects of this depleted background\cite{NESM4}.
This halo model accounts for the otherwise remarkable fact that galaxies
of all shapes are embedded in essentially spherical halos.
\par What, if any,  would be the gravitational effect of a depleted 
background density?  An analogy, in well-known physics, is vacancy
scattering of electrons in conductors.  In a complex material with
a regular periodic lattice independent electron waves are by no means 
trivial functions, but they propagate without contributing to scattering
or resistivity unless there is some lattice irregularity, 
such as a vacancy.  Impurity scattering depends on the difference
between impurity and host atomic T-matrices\cite{NES98}.  Similarly,
a photon or isolated mass particle follows a geodesic in the cosmic
background unless there is some disturbance of the uniform density 
$\rho_m$.  Both the condensed galactic density $\rho_g$ and the extended
subtracted density $-\rho_m$ must contribute to the deflection of 
background geodesics.  Such effects would be observed as gravitational 
lensing of photons and as radial acceleration of orbiting mass 
particles, following the basic concepts of general relativity.
\par Conformal analysis of galactic rotation, not assuming dark matter
\cite{MAN06,MAO11}, fits observed velocities consistent with empirical 
regularities.  Excessive centripetal radial acceleration independent 
of galactic mass is associated with an extragalactic source\cite{MAN97}.
The conformal Higgs model\cite{NESM1}, not invoking dark matter, infers 
positive (centrifugal) acceleration weight $\Omega_q$\cite{NESM1} due to
the cosmic background.  In the current epoch this is dominated by dark 
energy, due to the universal Higgs mechanism\cite{NESM2}, which is not 
affected by galaxy formation. In conformal theory, $\Omega_m$ is 
negative for positive mass because gravitational coefficient 
${\bar\kappa}$ is negative for a scalar field\cite{MAN06,NESM1}.  
Hubble weight $\Omega_m$, negative and currently small, contributes to 
positive acceleration $\Omega_q$.  Reduction of $\Omega_m$ by removal 
of mass in a depleted sphere implies a decrease of $\Omega_q$ relative 
to the cosmic background\cite{NESM4}. This is consistent with observed 
centripetal acceleration attributed to a galactic halo.
\par The effect of subtracted density $-\rho_m$ in standard Einsteinian
gravity would be centrifugal radial acceleration, contrary to what
is observed.  The challenge to $\Lambda$CDM is to incorporate 
or explain away the gravitational effect of missing matter of total 
mass $M$ that is drawn into an observed galaxy.  It seems unlikely that
a net mass $-M$ can simply be ignored.  A similar problem 
occurs for MOND, which postulates standard gravity, but scales the
implied acceleration by factor $\mu(a/a_0)$ without changing its sign.
\par Conformal gravity resolves this sign conflict in a fundamental but
quite idiosyncratic manner. Uniform, isotropic source density eliminates
the conformal Weyl tensor and its resulting gravitational effects.
In the conformal Higgs model, this leaves a modified gravitational
field equation due to the scalar field Lagrangian density.
The effective gravitational constant differs in sign and magnitude from
standard theory.  Hence the effect of a depleted halo should be
centripetal, as observed.  Analysis based on Newton-Einstein
constant $G$ is inappropriate for uniform, isotropic geometry,  
including the use of Planck-scale units for the early universe. 
\par The depleted halo model removes a particular conceptual problem
affecting analysis of anomalous galactic rotation in conformal gravity
theory\cite{MAN97,MAN06,MAO11}.  In empirical parameter 
$\gamma=\gamma^*N^*+\gamma_0$, $\gamma_0$ does not depend on
galactic mass, so must be due to the surrounding cosmos\cite{MAN97}.
Mannheim considers this to represent the net effect of distant matter,
integrated out to infinity\cite{MAN06}.  Divergent effects may not be a
problem, since external effects would be cut off by integration 
constants $\kappa$, as in recent fits to orbital velocity 
data\cite{MAO11}.  However, since the corresponding interior term, 
coefficient $\gamma^*N^*$, is centripetal, one might expect the exterior
term to describe attraction toward an exterior source, hence a net
centrifugal effect.  However, if coefficient $\gamma_0$ is due to a
subtractive halo, the implied sign change predicts net centripetal 
acceleration, in agreement with observation.
\par Integration parameter $\kappa$, included in fitting rotation
data\cite{MAO11}, acts to cut off gravitational acceleration at a
boundary radius.  In the halo model\cite{NESM4}, $\kappa$ is determined
by the boundary condition of continuous acceleration field at halo
radius $r_H$, determined by galactic mass, except for the nonclassical
linear potential term due to to the baryonic galactic core.
Three independent terms in effective gravitational potential $B(r)$,
each including a $\kappa r^2$ cutoff, contribute to orbital velocity
$v$\cite{NESM4}:
\begin{eqnarray}
v^2_{core}=\frac{GM}{r}(1-r^3/r_H^3),\\
v^2_{halo}=\half\gamma_0r(1-r/r_H),\\
\label{vext}
v^2_{ext}=\half N^*\gamma^*r(1-r/r_*),
\end{eqnarray}
for $r$ between galactic radius $r_g$ and halo radius $r_H$.
Parameters $\kappa_{core}=GM/r_H^3$ and $\kappa_{halo}=\gamma_0/2r_H$
cut off the acceleration field at $r_H$.
Geodesic deflection within halo radius $r_H$ is caused
by the difference between gravitational acceleration due to $\rho_g$
and that due to $\rho_m$\cite{NESM4}.
Because mass density $\Delta\rho=\rho_g-\rho_m$ inside $r_H$ integrates
to zero, the Keplerian core term terminates at $r_H$.
$\kappa^*=\gamma^*/2r_*$ may depend on galactic cluster environment.
\par The conformal gravitational field equation is 
\begin{eqnarray} \label{cfeq}
X_g^{\mu\nu}+X_\Phi^{\mu\nu}=\half\Theta_m^{\mu\nu},
\end{eqnarray}
which has an exact solution in the depleted halo, where 
$\Theta_m^{\mu\nu}=0$. Outside $r_g$, the source-free solution of
$X_g^{\mu\nu}=0$ in the ES metric\cite{MAK89} determines parameters 
proportional to galactic mass.  $X_\Phi^{\mu\nu}=0$ is solved in the RW 
metric as a modified Friedmann equation without $\rho_m$.  This
determines $\Omega_q(halo)$, which differs from $\Omega_q(cosmos)$
determined by $X_\Phi^{\mu\nu}=\half\Theta_m^{\mu\nu}(\rho_m)$.  These 
two equations establish a relation between $\Delta\Omega_q=
\Omega_q(halo)-\Omega_q(cosmos)$ and $\Delta\rho=\rho_g-\rho_m$,
which reduces to $-\rho_m$ in the halo.
\par Geodesic deflection in the halo is due to net gravitational 
acceleration $\Delta\Omega_q$, caused by $\Delta\rho$. Because the 
metric tensor is common to all three equations, the otherwise free 
parameter $\gamma_0$ in equation $X_g^{\mu\nu}=0$ must be compatible 
with the $X_\Phi$  equations.  This can be approximated in the halo 
(where $X_g^{\mu\nu}=0$) by solving equation
$\Delta X_\Phi^{\mu\nu}=\half\Delta\Theta_m^{\mu\nu}$, where
$\Delta X_\Phi= X_\Phi(halo)- X_\Phi(cosmos)$ and  
$\Delta\Theta_m=\Theta_m(halo)-\Theta_m(cosmos)=-\Theta_m(\rho_m)$.
Integration constant $\gamma_0$ is determined by conformal
transformation to the ES metric.  
\par From the modified Friedmann equation, acceleration weight
$\Omega_q(cosmos)=1-\Omega_\Lambda-\Omega_k-\Omega_m$.
Assuming a gravitationally flat true vacuum, $\Omega_m=0$ implies
$\Omega_k=0$ in the halo.  Equation $X_\Phi^{\mu\nu}(halo)=0$
implies $\Omega_q(halo)=1-\Omega_\Lambda$ , so that 
$\Delta\Omega_q=\Omega_k+\Omega_m$.
If $|\Delta\Omega_q|\ll\Omega_q$,
dark energy weight $\Omega_\Lambda$ cancels out, as does any vacuum  
value of $k$ independent of $\rho_m$.  Because both $\Omega_k$ and
$\Omega_m$ contain negative coefficients, if $\rho_m$ implies positive 
$k$ in the cosmic background, $\Delta\Omega_q$ is negative,
producing centripetal acceleration.  Thus positive $\gamma_0$, 
deduced from galactic rotation, is determined by the cosmic background,
as anticipated by Mannheim\cite{MAN06}.
\par To summarize the logic of the present derivation, Eq.(\ref{cfeq}) 
has an exact solution for $r_g\leq r\leq r_H$, outside the observable
galaxy but inside its halo, assumed to be a true vacuum with
$\Theta_m^{\mu\nu}=0$ because all matter has been absorbed into the 
galactic core.  For an isolated galaxy, 
$\Omega_q$ is nonzero, dominated at present time $t_0$
by dark energy $\Omega_\Lambda$.  Observed effects due to deflection
of background geodesics measure difference function
$\Delta\Omega_q=\Omega_k+\Omega_m$, inferred from the inhomogeneous
cosmic Friedmann equation in the RW metric.  Observable $\gamma_0$
is determined by transformation into the ES metric.
\par ES and RW metrics are related by a conformal transformation such
that $|k|=\frth\gamma_0^2$\cite{MAK89}, subject to analytic condition
$k\gamma_0<0$\cite{NESM4}. This relates solutions of the field 
equations. At present time $t_0$, with $a(t_0)=1$ and $h(t_0)=1$,
$\gamma_0^2=-4\Omega_k=-4\Delta\Omega_q$ in Hubble units $H_0^2/c^2$,
if $\Omega_m$ can be neglected.  Empirical coefficient
$\gamma_0=3.06\times 10^{-30}cm^{-1}$, deduced from anomalous galactic 
rotation velocities\cite{MAN97,MAO11}, implies 
$\Omega_k=-0.403\times 10^{-3}$, consistent with
consensus empirical value $\Omega_k=-0.002\pm0.011$\cite{KOM11}. 
\par The depleted conformal halo model implies that a galaxy of mass
$M$ produces a halo of exactly equal and opposite mass deficit.
Hence the ratio of radii $r_H/r_g$ should be very large, the cube root
of the mass-density ratio $\rho_g/\rho_m$.  Thus if the latter ratio is
$10^5$ a galaxy of radius 10kpc would be accompanied by a halo of radius
$10\times 10^{5/3}=464$kpc.  Equivalence of galactic and displaced halo
mass resolves the paradox for $\Lambda$CDM that despite any interaction
other than gravity, the amount of dark matter inferred for a galactic
halo is strongly correlated with the galactic luminosity or baryonic
mass\cite{SAL07,MCG05}.
The skew-tensor theory of Moffat\cite{BAM06}
avoids this problem by an additional long-range field generated by
the baryonic galaxy.  Renormalization group flow of parameters
models the MOND postulate of modifiied Newtonian acceleration.

\section{Open questions}
\par The conformal Higgs model accounts for scalar field parameter
$w^2$, which becomes universal dark energy.  Conformal symmetry does
not preclude nonzero $\lambda$ in the bare scalar field Lagrangian 
density.  The possible time dependence of $\lambda$ is not known.
Empirical value $\lambda=-0.189\times 10^{-88}$ follows from
well-established values of Hubble dark-energy weight $\Omega_\Lambda$
and Higgs scalar field amplitude $\phi_0$, but is not determined by
theory limited to neutral gauge field $Z_\mu$.  Analysis of the coupled 
field equations incorporating charged gauge fields $W^\pm_\mu$ involves 
conceptual difficulties, not yet resolved, regarding self-interaction
and an electrically charged vacuum.
\par Although conformal theory implies an initial epoch of rapid,
inflationary Hubble expansion, this cannot be treated in detail until
the time-dependence of several nominally constant parameters is known.
The modified Friedmann equation determines the time variation of Ricci
scalar $R$ on a cosmological scale 
(Hubble time unit $1/H_0=4.38\times 10^{17}s$).  
Implied rate scale ${\dot\phi}_0/\phi_0$ affects other parameters.
Whether or not conformal theory can explain empirical data relevant to
the "big-bang" model, such as relative deuterium abundance
and nucleosynthesis in general, cannot be tested until the early
time dependence of parameters is known.
\par The conformal halo model apparently eliminates the need for dark
matter for an isolated galaxy.  The implications for galactic clusters
have not been explored.  Individual halo mass is only part of the 
dark matter inventory for clusters\cite{AEM11}.  The conformal
long-range interaction between galaxies whose halos do not overlap
determines Eq.(\ref{vext}).  Analysis of the implications for a
galactic cluster has not yet been carried out.  
\par Crucially, the classical Newtonian virial theorem is not valid,
so observed high thermal energy within a galactic cluster cannot be
used without entirely new dynamical analysis to estimate the balance
between baryonic matter, radiative energy, and hypothetical dark matter.
These remarks apply directly to models of galaxy formation.
In the conformal halo model, any growing galaxy is stabilized by the
net gravitational effect of its accompanying depleted halo.
A detailed dynamical model has not yet been worked out. 

\section{Conclusions}
\par Conformal theory can explain the existence of galactic halos and 
the existence and magnitude of dark energy.  Cosmological data
including anomalous galactic rotation velocities and parameters 
relevant to Hubble expansion are fitted without invoking dark matter.
Conformal gravity, the conformal Higgs model, and the depleted 
halo model are mutually consistent, removing several paradoxes or
apparent logical contradictions in cosmology.  
\par In uniform, isotropic (Robertson-Walker) geometry, the Weyl
tensor basic to conformal gravity vanishes identically.  Observed
gravitational acceleration can be attributed to a background 
scalar field, identified here with a conformal Higgs field.
The implied Hubble
expansion agrees with supernovae redshift data and determines
centrifugal acceleration in the early universe, as required
for a spontaneous big-bang model.  The tachyonic mass parameter
in the conformal Higgs model is identified with dark energy, which is
simply a secondary consequence of the SU(2) symmetry-breaking finite
scalar field amplitude required to explain weak gauge boson masses. 
This tachyonic mass parameter is generated by a new and very small
scale parameter, the cosmological time derivative of the gravitational
Ricci scalar.  This removes a longstanding apparent conflict between
magnitudes of elementary-particle and cosmological parameters.


\begin{thebibliography}{99}
\bibitem{DOD03} Dodelson, S. 
{\em Modern Cosmology};
Academic Press: New York, USA, 2003.
\bibitem{WEY18} Weyl, H.
Gravitation und Elektrizit\"at.
{\em Sitzungber.Preuss.Akad.Wiss.} {\bf 1918}, {}, 465-480.
\bibitem{MAN90} Mannheim, P.D.
Conformal Cosmology with No Cosmological Constant.
{\em Gen.Rel.Grav.} {\bf 1990}, {\em 22}, 289-298.
\bibitem{MAN06} Mannheim, P.D.
Alternatives to Dark Matter and Dark Energy.
{\em Prog.Part.Nucl.Phys.} {\bf 2006}, {\em 56}, 340-445.
\bibitem{NESM1} Nesbet, R.K.
Cosmological Implications of Conformal Field Theory.
{\em Mod.Phys.Lett.A} {\bf 2011}, {\em 26}, 893-900.
\bibitem{DEW64} DeWitt, B.S., in
{\em Relativity, Groups, and Topology};
DeWitt, C.; DeWitt, B.S., eds.;
Gordon and Breach: New York, USA, 1964.
\bibitem{MAN11} Mannheim. P.D.
Comprehensive Solution to the Cosmological Constant, Zero-Point
Energy, and Quantum Gravity Problems.
{\em Gen.Rel.Grav.} {\bf 2011}, {\em 43}, 703-750. 
\bibitem{NESM2} Nesbet, R.K.
Conformal Higgs Model of Dark Energy.
(arXiv:1004.5097v2 [physics.gen-ph]).
\bibitem{MAK89} Mannheim, P.D.; Kazanas, D.
Exact Vacuum Solution to Conformal Weyl Gravity and Galactic 
Rotation Curves.
{\em Astrophys.J.} {\bf 1989}, {\em 342}, 635-638.
\bibitem{FLA06} Flanagan, E.E.
Fourth Order Weyl Gravity.
{\em Phys.Rev.D} {\bf 2006}, {\em 74}, 023002.
\bibitem{MAN07} Mannheim, P.D.
Schwarzschild Limit of Conformal Gravity in the Presence of Macroscopic
Scalar Fields.
{\em Phys.Rev.D} {\bf 2007}, {\em 75}, 124006.
\bibitem{MAN97} Mannheim, P.D.
Are Galactic Rotation Curves Really Flat?
{\em Astrophys.J.} {\bf 1997}, {\em 479}, 659-664.
\bibitem{MAO11} Mannheim, P.D.; O'Brien, J.G. 
Impact of a Global Quadratic Potential on Galactic Rotation Curves.
{\em Phys.Rev.Lett.} {\bf 2011}, {\em 106}, 121101.
\bibitem{MAO12} Mannheim, P.D.; O'Brien, J.G.
Fitting Galactic Rotation Curves with Conformal Gravity and a Global
Quadratic Potential.
{\em Phys.Rev.D} {\bf 2012}, {\em 85}, 124020.
\bibitem{OAM12} O'Brien, J.G.; Mannheim, P.D.
Fitting Dwarf Galaxy Rotation Curves with Conformal Gravity.
{\em Mon.Not.R.Astron.Soc.} {\bf 2012}, {\em 421}, 1273-1282.
\bibitem{NESM4} Nesbet, R.K. 
Proposed Explanation of Galactic Halos.
(arXiv:1109.3626v3 [physics.gen-ph]).
\bibitem{PAS95} Peskin, M.E.: Schroeder, D.V.
{\em Introduction to Quantum Field Theory};
Westview Press: Boulder, CO, 1995.
\bibitem{CAG98} Cottingham, W.N.; Greenwood, D.A.
{\em An Introduction to the Standard Model of Particle Physics};
Cambridge Univ. Press: New York, 1998.
\bibitem{NES03} Nesbet, R.K. 
{\em Variational Principles and Methods in
Theoretical Physics and Chemistry};
Cambridge Univ. Press: NY, 2003.
\bibitem{DAF05} Deser, S.; Franklin, J.  
Schwarzschild and Birkhoff a la Weyl.
{\em Am.J.Phys.} {\bf 2005}, {\em 73}, 261-264.
\bibitem{OAP73} Ostriker, J.P.; Peebles, P.J.E.
A Numerical Study of the Stability of Flattened Galaxies: or, Can
Cold Galaxies Survive?
{\em Astrophys.J.} {\bf 1973}, {\em 186}, 467-480.
\bibitem{OPY74} Ostriker, J.P.; Peebles, P.J.E.; Yahil, A.
The Size and Mass of Galaxies and the Mass of the Universe.
{\em Astrophys.J.} {\bf 1974}, {\em 193}, L1-L4.
\bibitem{SAN10} Sanders, R.H. {\em The Dark Matter Problem};
Cambridge Univ. Press: NY, 2010.
\bibitem{RAS72} Rogstad, D.H.; Shostak, G.S.
Gross Properties of Five SCD Galaxies.
{\em Astrophys.J.} {\bf 1972}, {\em 176}, 315-321. 
\bibitem{BOS81} Bosma, A.
21-cm Line Studies of Spiral Galaxies.
{\em Astron.J.} {\bf 1981}, {\em 86}, 1825-1846. 
\bibitem{WCW79} Walsh, D.; Carswell, R. F; Weymann, R. J.
Twin Quasistellar Objects or Gravitational Lens.
{\em Nature} {\bf 1979}, {\em 279}, 381-384.
\bibitem{PAC87} Paczynski, B.
Giant Luminous Arcs Discovered in Two Clusters of Galaxies.
{\em Nature} {\bf 1987}, {\em 325}, 572.
\bibitem{BAR10} Bartelmann, M.
Gravitational Lensing.
{\em Classical and Quantum Gravity} {\bf 2011} {\em 27}, 233001.
\bibitem{PSS98} Persic, M.; Salucci, P.; Stel, F.
The Universal Rotation Curve of Spiral Galaxies. I. 
{\em Mon.Not.R.Astron.Soc.} {\bf 1996}, {\em 281}, 27-48.
\bibitem{SAL07} Salucci, P. et al.
The Universal Rotation Curve of Spiral Galaxies. II. 
{\em Mon.Not.R.Astron.Soc.} {\bf 2007}, {\em 378}, 41-51.
\bibitem{MIL83} Milgrom, M.
A Modification of Newtonian Dynamics.
{\em Astrophys.J.} {\bf 1983}, {\em 270}, 365-370. 
\bibitem{SAM02} Sanders, R.H.; McGaugh, S.S.
Modified Newtonian Dynamics as an Alternative to Dark Matter.
{\em Ann.Rev.Astron.Astrophys.} {\bf 2002}, {\em 40}, 263-317.
\bibitem{TAF77} Tully, R.B.; Fisher, J.R.
A New Method for Determining the Distances to Galaxies.
{\em Astron.Astrophys.} {\bf 1977}, {\em 54}, 661-673. 
\bibitem{MCG11} McGaugh, S.S.
Novel Test of MOND with Gas Rich Galaxies. 
{\em Phys.Rev.Lett.} {\bf 2011}, {\em 106}, 121303.
\bibitem{BEK04} Bekenstein, J.D.
Relativistic Gravitation Theory for the MOND Paradigm.
{\em Phys.Rev.D} {\bf 2004}, {\em 70}, 083509.
\bibitem{MOF06} Moffat, J.W.
Scalar-Tensor-Vector Gravity Theory
{\em JCAP} {\bf 2006}, {\em 2006},4-23 .
\bibitem{BAM06} Brownstein, J.R.; Moffat, J.W.
Galaxy Rotation Curves without Non-Baryonic Dark Matter.
{\em Astrophys.J.} {\bf 2006}, {\em 636}, 721-761.
\bibitem{MRT12} Moffat, J.W.; Rahvar,S.; Toth, V.T.
Applying MOG to Lensing: Einstein Rings, Abell 520 and the Bullet
Cluster. (arXiv:1204.2985v1[astro-ph.CO]).
\bibitem{HUB29} Hubble, E.
A Relation Between Distance and Radial Velocity among Extra-Galactic 
Nebulae.
{\em Proc.Nat.Acad.Sci.} {\bf 1929}, {\em 15}.
\bibitem{RAS98} Riess, A.G. et al.
Observational Evidence from Supernovae for an Accelerating Universe.
{\em Astron.J.} {\bf 1998}, {\em 116}, 1009-1038.
\bibitem{PER99} Perlmutter, S. et al.
Measurement of $\Omega$ and $\Lambda$ from 42 High-redshift Supernovae.
{\em Astrophys.J.} {\bf 1999}, {\em 517}, 565-586.
\bibitem{FRI22} Friedmann, A.
\"Uber die Kr\"ummung des Raumes.
{\em Zeits.Phys.} {\bf 1922}, {\em 10}, 377-386.
\bibitem{FRI24} Friedmann, A.
\"Uber die M\"oglichkeit einer Welt mit konstanter negativer Kr\"ummung
des Raumes.
{\em Zeits.Phys.} {\bf 1924}, {\em 21}, 326-332.
\bibitem{LEM27} LeMa\^itre, G.
Un Univers Homog\`ene de Masse Constante et de Rayon Croissant.
{\em Ann.Soc.Sci.Bruxelles} {\bf 1927}, {\em A47}, 49-59.
\bibitem{MAN03} Mannheim, P.D.
How Recent is Cosmic Acceleration?
{\em Int.J.Mod.Phys.D} {\bf 2003}, {\em 12}, 893-904.
\bibitem{KOM09} Komatsu, E. et al.
Five-year WMAP Observations. 
{\em Astrophys.J.Supp.} {\bf 2009}, {\em 180}, 330-381.
\bibitem{KOM11} Komatsu, E. et al.
Seven-year WMAP Observations. 
{\em Astrophys.J.Supp.} {\bf 2011}, {\em 192}, 18-74.
\bibitem{WAM07} Wang, Y.; Mukherjee, P.
Observational Constraints on Dark Energy and Cosmic Curvature.
{\em Phys.Rev.D} {\bf 2007}, {\em 76}, 103533.
\bibitem{NESM3} Nesbet, R.K. 
The Higgs Scalar Field with no Massive Higgs Particle. 
(arXiv:1009.1372v3 [physics.gen-ph]).
\bibitem{CHO12} Cho, A.
Higgs Boson Makes its Debut after Decades-Long Search.
{\em Science} {\bf 2012}, {\em 337}, 141-143. 
\bibitem{NES98} Nesbet, R.K.
Theory of Spin-dependent Conductivity in GMR Materials.
{\em IBM J.Res.Develop.} {\bf 1998}, {\em 42}, 53-71.
\bibitem{MCG05} McGaugh, S.S.
Balance of Dark and Luminous Mass in Rotating Galaxies.
{\em Phys.Rev.Lett.} {\bf 2005}, {\em 95}, 171302.
\bibitem{AEM11} Allen, S.W.; Evrard, A.E.; Mantz, A.B.
Cosmological Parameters from Observations of Galaxy Clusters.
{\em ARA\&A} {\bf 2011}, {\em 49}, 409-470. 
\end{thebibliography}
\end{document}